\documentclass[]{aastex}

\def\spose#1{\hbox to 0pt{#1\hss}} 

\def\simlt{\mathrel{\spose{\lower 3pt\hbox{$\mathchar"218$}}      

\raise 2.0pt\hbox{$\mathchar"13C$}}} 

\def\simgt{\mathrel{\spose{\lower 3pt\hbox{$\mathchar"218$}}      

\raise 2.0pt\hbox{$\mathchar"13E$}}} 

\def\eg{{\rm e.g.~}} \def\ie{{\rm i.e.~}} 

\def\etal{{\rm et~al.~}} 

\shorttitle{Missing in elliptical galaxies} 

\begin{document}

\title{What are we missing in elliptical galaxies ? }

\author{Jeremy Mould\footnote{ARC Centre of Excellence for All-sky Astrophysics}} 
\affil{Centre for Astrophysics and Supercomputing, Swinburne University, Vic 3122, Australia} 
\authoraddr{E-mail: jmould@swin.edu.au} 


\begin{abstract}
The scaling relation for early type galaxies in the 6dF galaxy  survey
does not have the velocity dispersion dependence expected from
standard stellar population models. As noted in recent work with SDSS, there seems to be an additional dependence
of mass to light ratio with velocity dispersion, possibly due to a bottom heavy initial mass function. Here
we offer a new understanding of the 6dF galaxy survey 3D gaussian Fundamental Plane in terms of a
parameterized Jeans equation, but leave mass dependence
of M/L and mass dependence of structure still degenerate with just
the present constraints. Hybrid models have been proposed recently.
Our new analysis brings into focus promising lines of enquiry which could be pursued to lift this degeneracy, including stellar atmospheres computation, kinematic probes of ellipticals at large radius, and a large sample of one micron spectra.

\end{abstract}

\section{Introduction}

Scaling relations for galaxies, like the Tully-Fisher relation for disks
and the Faber-Jackson relation for ellipticals, are fundamental and powerful.
They challenge theories of galaxy formation and they allow us to measure galaxy
distances. Local galaxy distances provide us with maps of the mass distribution
to compare with the light distribution.

\cite{TF77} and \cite{FJ} pioneered scaling relations,
and the latter was soon replaced with the fundamental plane \citep{L87}, or FP. The virial theorem offered a partial explanation, \eg \cite{AM79,F87}. 
Nearly two decades ago \cite{W96} opined that an understanding of scaling relations was within reach.
But hydrodynamic models of galaxy formation
and semi-analytic models have not led to a full understanding.

Nor has the comparison of mass maps and the galaxy distribution yet
led to a satisfying resolution. On the one hand, the distribution of peculiar
velocities in x-ray clusters of the 6dF galaxy  survey (Magoulas 2012; PhD thesis\footnote{http://dtl.unimelb.edu.au} )
and of Planck's kinetic Sunyaev Zeldovich peculiar velocities \citep{A13} is comparable;
on the other, the bulk flow measured locally \citep{F10, M12} is on the high side of expectations from the $\Lambda$CDM model.

In this presentation we outline the scaling relation problem for early type galaxies
from the perspective of the 6dF galaxy  survey \citep{J05, LC}.
Our findings parallel the SDSS result of \cite{C13}. We consider
what this means for the elliptical scaling relation. And we suggest what needs
to be done to test the notion that a varying bottom-heavy initial mass function (IMF)
is what we are missing in ellipticals.
 
\section{Mass to light ratio}

\cite{M12} fitted a 3D Gaussian to the surface brightnesses, radii,
and velocity dispersions of over 11000 early type galaxies in the 6dF 
survey. These variables are related to the dynamical mass by the virial theorem in appropriate units\footnote{Solar units and AU
with velocities measured in 30 km s$^{-1}$ units.}

$$M_{total} = k \sigma^2 r_e \eqno (1) $$

and

$$L = 2  \pi I_e r_e^2 \eqno (2)$$

The numerical constant $k$ depends on the galactic structure, \eg it is $(\surd 2 - 1)^{-1}$ for a Hernquist profile \citep{H}.
\cite{SC12} and \cite{P08} measured Lick indices for the galaxies, giving estimates
of metallicity, $Z$, and age, $t$, for their stellar populations.
\cite{M07} models\footnote{
http://www.icg.port.ac.uk/$\sim$maraston/Claudia\%27s\_Stellar\_Population\_Model.html}
 for the 2MASS J band \citep{2MASS} adopted in the Magoulas dataset predict:

$$M_{model}/L = M/L(Z,t) \eqno (3)$$

for some assumed IMF and horizontal branch distribution.

The foregoing equations allow us to plot Figure 1.

\begin{figure}[ht]
\begin{center}
\includegraphics[clip, angle=-90, width=.9\textwidth]{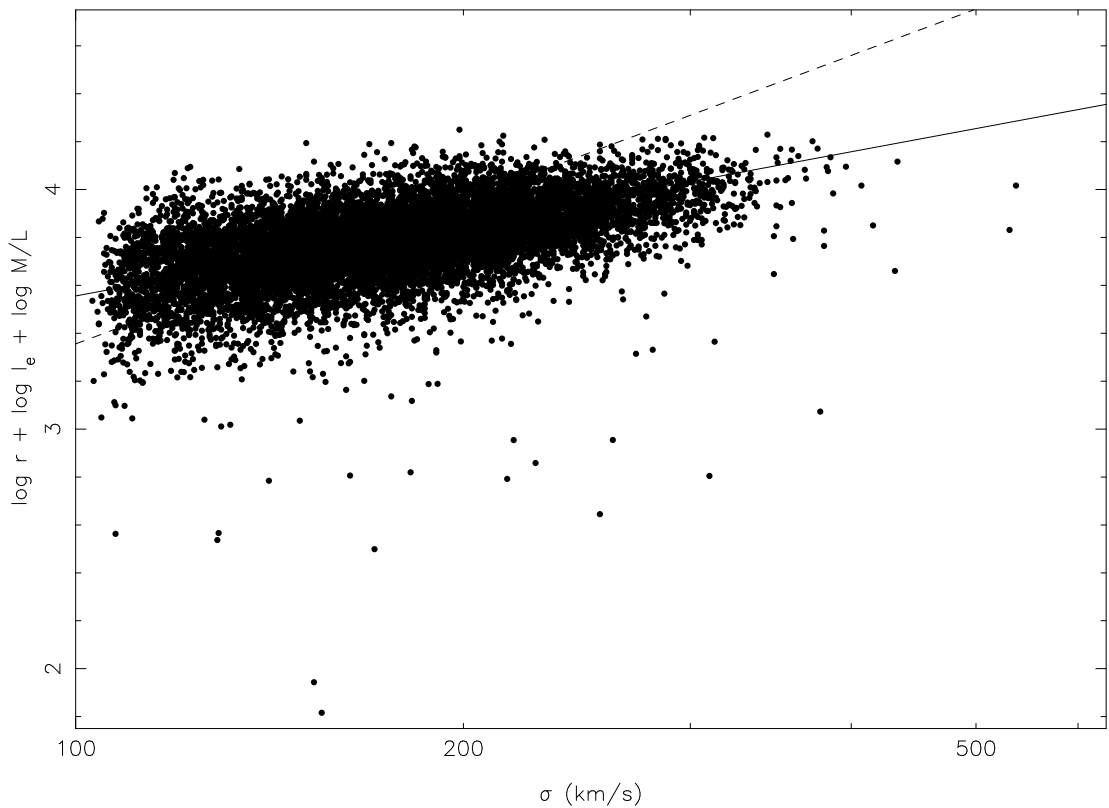}
\end{center}
\caption{Model mass to light ratio combined with surface brightness and radius versus
velocity dispersion. The dashed line is equation (4); the solid line is equation (5). Fits are weighted by the observational errors. 
The scatter in the ordinate is 0.13 dex.}
\end{figure}

If $M_{total}~ = ~M_{model} ~=~M$, we expect

$$ 2 \pi I_e r_e M/L = k \sigma^2 \eqno(4)$$

However, what we observe is

$$I_e r_e M/L \propto \sigma \eqno(5)$$

This implies that the hypothesis is not correct, and that rather

$$M_{total}/M_{model} \propto \sigma \eqno (6)$$.

\cite{C13} see something similar in SDSS. This prompts the question
which is the title of this Letter. In ATLAS$^{3D}$ \cite{Ca} find M/L $\sim~\sigma_e^{0.72}$
where $\sigma_e$ is the velocity dispersion within the effective radius,
and this is consistent with equation~(5). By making dynamical models, they affirm
that the effective radius in equation (1) is the same as the effective radius in equation (2).
Our radii are calculated from the galaxy redshifts neglecting peculiar velocities.
If the $rms$ peculiar velocity is $\sim$200--500 km s$^{-1}$ \citep{P76, P87}, the corresponding error in the
ordinate of Figure 1 is 2--5\% for a galaxy at 10$^4$  km s$^{-1}$ and is not a significant contributor to the scatter.
The majority of the sample is at higher redshift, \ie in the Hubble flow for present purposes.
\cite{A09} saw a similar dependence of M/L on $\sigma$, fitting stellar population models to galaxies in the Coma cluster.

A possible concern is selection effects in the 6dF Galaxy Survey. Could we be missing a cloud of low $I_e$, low $r_e$ galaxies at $\sigma~\approx$ 100 km s$^{-1}$ ?
\cite{M12} have calculated selection probabilities for the sample and these are implemented in a V/V$_{max}$ manner.
When this correction is turned on, the fitted logarithmic slope in Figure 1 rises from 0.79 to 1.00. Sample selection is therefore not the source of the discrepancy
between equations 5 \& 6.

Much discussion has centered around representations of the variables involved as a tilted plane in a data cube. Since we have five variables and three equations in the present analysis, the FP
 ~approach may have most merit as a diagnostic tool for checking the three equations against the data. While the FP in 3-space is preferable to a  volume in the full 5 variable hypercube, the approach we take here still allows us to open up in the next two sections the important physical questions. These relate more to parameters which have not yet been measured than to those five that have.

\section{The initial mass function}

Missing mass is usually interpreted as dark matter. But it would be naive to immediately assume this is the
basis for equation (6) for two reasons. Our measurements are made
inside the half-light radii of early type galaxies. We believe these are
baryon dominated. Dark matter only dominates well beyond 10 kpc in large galaxies. Second, it is the very smallest dwarf elliptical galaxies where
mass to light ratios reach 100 and dark matter is the primary constituent.
Figure 1 is for galaxies with $\sigma~>~100$ km s$^{-1}$, the cutoff of the 6dF
spectrograph.

And so, like \cite{C13}, we are led to a bottom-heavy IMF hypothesis
to explain equation (6). This hypothesis is an old one. \cite{ST71} conjectured that the strong lined M31, M32 and M81 nuclei might have
power law IMFs $n(m) ~\propto~ m^{-s}$ with $s ~>>$ 2.35, the Salpeter value.
The 0.8--2 micron spectra of these galaxies would be dominated by M dwarfs,
rather than the conventional giant branch.

One micron spectra of nearby ellipticals, however, have not led to clear
confirmation. One the one hand, \cite{C12}, \cite{S12},
\cite{F13}, and \cite{S13} favor a Salpeter IMF. On
the other, there is no clear trend in their fitted IMF with velocity dispersion. 
\cite{Cb} find a transition of the mean IMF from Kroupa to Salpeter in the interval 
$\sigma_e~\simeq$ 90--290 ~km~ s$^{-1}$, with a smooth variation in between. \cite{D13} find
from mass models, not stellar population models,
a mass-dependent IMF which is {\it lighter} than Salpeter at low masses 
and {\it heavier} than Salpeter at high masses. 

Assuming power law IMFs persist to Jupiter masses, M/L is very sensitive to $s$, as shown in Figure 2.  
In this case, the difference between equations 5 \& 6 can be accommodated by a modest dependence of $s$ on $\sigma$.

\begin{figure}[ht]
\begin{center}
\includegraphics[clip, angle=-90, width=0.4\textwidth]{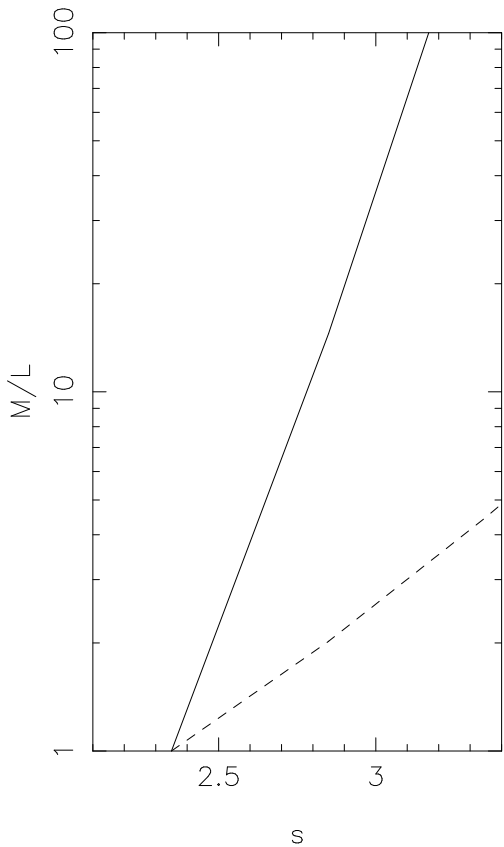}
\end{center}
\caption{The mass to light ratio relative to that from a Salpeter IMF, assuming that a power law IMF persists to Jupiter masses.
White dwarfs of 0.6M$_\odot$ are assumed to be the progeny of stars of initial mass 0.8--8M$_\odot$.
In this $s$-regime the effect of higher mass remnants is negligible. The dependence is less steep if the lower mass cutoff is 10 Jupiters (dashed line).}
\end{figure}

\section{Galaxy structure}
Notwithstanding the similarity of dynamical and luminous radii 
found by \cite{Ca}, one can imagine a large elliptical
with a tidal radius of 10 kpc surrounded by a 100 kpc spherical shell of dark matter of equal mass.
Its mass half radius (equation~1) would be much larger than its luminosity half radius (equation 2)
and it could lie on the dashed line of Figure 1. 
Is the baryonic fraction of high $\sigma$ ellipticals smaller
than that of low $\sigma$ ellipticals due to structural differences 
related to the fraction of the dark halo that is occupied by baryons?
At first blush this idea does not sit well with the observation
that it is ellipticals like the Draco dwarf galaxy that have very low
baryonic fractions \citep{M98}, rather than giants like M49 in the Virgo
cluster. However, there may be not one, but two, factors at work.
Almost all theories of dwarf elliptical galaxies' low metallicity,
\eg \cite{L74, K11, M84},    
involve loss of a major part of the primordial gas due to feedback, probably
simply the action of the first supernovae to explode in these
shallow gravitational potentials. One class of solutions to the missing satellite problem
involves this or similar baryonic processes \citep{N}.
The mass metallicity relation is clearly delineated in our 6dF data \citep{SC12}.

So, two separate phenomena may be required to account for the baryonic fraction
in ellipticals, (1) an initial underfilling of the largest halos with baryons,
and (2) feedback evacuating the smallest halos of baryons.
The kinematics of ellipticals' globular cluster systems
have been extended beyond~
6$r_e$ \citep{N12, F11, M90}, so these ideas can be tested.

A more quantitative model can be constructed from the Jeans equation,

$$GM(r) = -r\sigma_{rr}^2 [~{\rm{d~ln}~\nu\over \rm{d~ln}{\it~r}} + {\rm{d~ln}~ \sigma_{rr}^2\over \rm{d~ln}{\it ~r}} + \beta(r)] \eqno(7)$$

In a potential with a flat circular velocity and isotropic  orbits with radial velocity dispersion $\sigma_{rr}$, only the first 
of the three terms in the brackets is nonzero. 
For a power law density distribution this is the index of the power law. If we manipulate this term so that the index is --4
for galaxies at the left of Figure 1 and flatter for larger velocity dispersions, we
can set the first term to 400(km s$^{-1}$)/$\sigma_{rr}$ and obtain in the units
of equation (1),

$$M_{total} =13.3~ r~\sigma_{rr}\eqno(8)$$

If this replaces equation (1), we obtain the same $ \sigma $ dependence in Figure 1 as equation (5). If dln$\nu$/dlnr 
= --4 with $\sigma$ = 100 km s$^{-1}$, the density distribution is the black curve in Figure 3. And if dln$\nu$/dlnr 
= --2 with $\sigma$ = 200 km s$^{-1}$, the density distribution is the green curve in Figure 3. The ratio of mass half radii between the red and black curves is 
between 3 and 4, depending on the precise treatment of the core. The red curve is closest to the expectation from CDM simulations \citep{NFW}.
This is a  rather extreme model compared with the  modest mass dependence of halo structure seen in CDM simulations by \cite{L13} and \cite{D14} with baryonic feedback.

\begin{figure}[ht]
\begin{center}
\includegraphics[clip, angle=-90, width=0.4\textwidth]{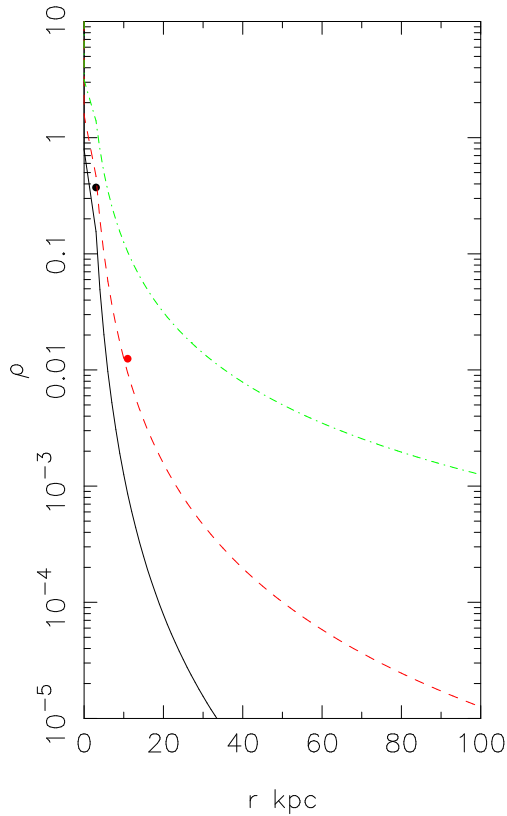}
\end{center}
\caption{Power law spherical density distributions with a core inside 1 kpc. The power law index is --2,--3,--4 in the  green (dotted), red (dashed), and black (solid)  curves respectively. Half mass radii in the latter two cases are marked. The NFW \citep{NFW} profile resembles the red curve at large radii and
the isothermal sphere resembles the green curve. }
\end{figure}

\section{Hybrid Models}

Although the structural model of equation (8) may be extreme and the other two terms in the brackets of equation (7) are also likely to
be closer to zero than one, we can parameterize departures from 
equations (1) \& (8) as

$$M \propto r \sigma^{1+\delta}\eqno(9)$$

with $0~<\delta~<1$. 

Our fit to the FP \citep{M12} can be written, together with equation (2) as

$$L \propto r_e^{0.9} \sigma^{1.65}\eqno(10)$$

Dividing the two at $r_e$ gives

$$M/L \propto r_e^{0.1} \sigma^{1+\delta-1.65}\eqno(11)$$

To the extent that equation (11) is almost independent of $r_e$, the pure structural approach to the elliptical galaxy scaling ratio
requires $\delta$ = 0.65.

But if we also parameterize M/L as $\sigma^\epsilon$, we have

$$\epsilon = \delta - 0.65\eqno(12)$$

This gives rise to the notion of hybrid models  \citep{O13}.
In other words, the scaling relation problem is currently
underconstrained. One can either propose a nonzero $\epsilon$
or a nonzero value of ($\delta$ -- 0.65), or both in the hybrid case. 

The Jeans equation tells us that, if ellipticals are formed from dry mergers, 
and their potentials in the outer parts are always close to NFW potentials, then $\delta$ can differ
from unity only in as much as dln$\sigma$/dlnr or $\beta$ are negative
and a function of $\sigma$ (\ie galaxy mass). If ellipticals form
in wet mergers, or if AGN feedback or adiabatic contraction are important, the baryons may force exceptions to NFW profiles, 
and $\delta$ is more free (until future hydrodynamic models constrain it).

A number of questions arise from this.

(1) Since there is a rising metallicity trend with velocity dispersion
in ellipticals, \eg \cite{SC12, GF}, is the IMF governed by metallicity ?

(2) Are the spectroscopic indicators of M dwarf enrichment metallicity
sensitive ? Hydride bands are strong in metal poor halo M dwarfs \citep{M76, B03}.
FeH is a hydride band with similar properties to CaH in molecular equilibria \citep{M78}.

(3) Is M dwarf enrichment independent of metallicity or age but directly
coupled to velocity dispersion through, possibly, a Jeans mass dependence
on the density of the collapsed and star-forming protogalaxy ? Two stages of star formation
are considered by \cite{W}.

(4) Are brown dwarfs enriched in high-$\sigma$ ellipticals, long faded from their L and T dwarf origins ?
The IMF does not end at the hydrogen burning limit \citep{C03, K12}.
Only lensing will find such objects \citep{B13, A93}. A heavyweight IMF is strongly disfavored
for the closest lensed early type galaxy \citep{SL}.

These questions in turn suggest further work to isolate what we are missing in ellipticals.

(1) The one micron spectra of stars and galaxies are eminently able to be
modeled and metallicity dependences predicted. The FeH band may offer some
challenges, but lines such as K I are straightforward, and the continuum
is well defined at high resolution, \eg \cite{M+07}.

(2) Early type galaxy redshift surveys (\eg TAIPAN\footnote{http://physics.mq.edu.au/astronomy/workshop\_2012/}) could be extended to one micron
to permit a principal components analysis separation of $Z$, $t$, and $s$.

(3) Hydrodynamic simulations of the formation of ellipticals are needed
right down to the star formation scale, so that theory can make 
a statement about the $0.5 M_\odot/0.1M_\odot$ stellar mass ratio expectations
for massive and intermediate mass ellipticals.
Analytic approaches look promising, e.g. \cite{Ho}.

(4) The dynamics of ellipticals must be probed to larger halo radii.

\section{Conclusions}

The ratio of dynamical mass to standard stellar population mass in the 6dF galaxy sample of
early type galaxies  is approximately proportional to velocity dispersion
in the range 100 $<~\sigma~<$ 300 km s$^{-1}$.

A bottom heavy IMF is a simpler explanation of this trend than the notion, for example,
that the baryonic fraction of these galaxies has a peak at $\sim$ 200 km s$^{-1}$ 
velocity dispersion. 
However, a greater dark matter mass fraction in large halo potentials is
an alternative hypothesis that cannot at present be ruled out. Hybrid models are very possible \citep{O13}.
Stellar atmospheres \citep {A97, AF13}, stellar populations, globular cluster dynamics, and galaxy formation theory can
all play valuable roles in tying down what we are missing in elliptical galaxies.

The question posed in the title of this Letter deserves a simple answer.
Our answer is that we are missing $\delta$ and $\epsilon$.
Unambiguous determination of the level of late M dwarf light in ellipticals will measure $\epsilon$. Kinematic probes of the outer
gravitational potentials of ellipticals will measure $\delta$,
as neutral hydrogen did for disk galaxies back when scaling relations
were first proposed.

\acknowledgements
We are grateful for ARC grants DP1092666 \& LP130100286 in partial support of this work. I am grateful to the ARC Centre of Excellence for Particle Physics
at the Terascale for the opportunity to present this work and to Darren Croton
for reading over the manuscript. 
Survey astronomy such as TAIPAN is supported by the ARC through CAASTRO{\footnote{www.caastro.org}. We thank the whole 6dF team for their contributions. The 6dF galaxy survey has been a project of the Australian Astronomical Observatory with the UK Schmidt Telescope. Data are archived at the Widefield Astronomy Unit of the Royal Observatory Edinburgh. This study has also used 2MASS data provided by NASA-JPL.




\begin{thebibliography}{}
\bibitem[Aaronson, Mould \& Huchra (1979)]{AM79}Aaronson, M., Mould, J. \& Huchra, J. 1979, ApJ 229, 1
\bibitem[Ade \etal (2013)]{A13}Ade, P. \etal 2013, astro-ph 1303.5090
\bibitem[Alcock \etal (1993)]{A93}Alcock, C. \etal 1993, Nature, 365, 621
\bibitem[Allanson \etal (2009)]{A09}Allanson, S. \etal 2009, ApJ, 702, 1275	
\bibitem[Allard \etal (1997)]{A97}Allard, F. \etal 1997, ARA\&A, 35, 137
\bibitem[Allard \etal (2013)]{AF13}Allard, F. \etal 2013, 
Memorie della Societa Astronomica Italiana Supplement, v.24, p.128  
\bibitem[Barnab\'{e} \etal (2013)]{B13}Barnab\'{e}, M. \etal 2013, arXiv 1306.2635
\bibitem[Burgasser \etal (2003)]{B03}Burgasser, A. \etal 2003, ApJ, 592, 1186
\bibitem[Campbell \etal (2014)]{LC}Campbell, L. \etal 2014, submitted to MNRAS
\bibitem[Cappellari \etal (2013a)]{Ca} Cappellari, M. \etal 2013a, MNRAS, 432, 1709
\bibitem[Cappellari \etal (2013b)]{Cb} Cappellari, M. \etal 2013b,	MNRAS, 432, 1862
\bibitem[Chabrier (2003)]{C03}Chabrier, G. 2003, PASP, 115, 763
\bibitem[Conroy \& van Dokkum (2012)]{C12}Conroy, C. \& van Dokkum, P. 2012, ApJ, 760, 71 
\bibitem[Conroy \etal (2013)]{C13}Conroy, C. \etal 2013, ApJL, in press, astro-ph 1306.2316
\bibitem[Di Cintio \etal (2014)]{D14}Dutton, A. 2014, MNRAS, 437, 415	
\bibitem[D'Onofrio \etal (2013)]{O13}D'Onofrio, M. \etal 2013, MNRAS, 435, 45
\bibitem[Dutton \etal (2013)]{D13}Dutton, A. 2013, MNRAS, 432, 2496	
\bibitem[Faber  \& Jackson (1976)]{FJ}Faber, S. \& Jackson, R. 1976, ApJ, 204, 668 
\bibitem[Faber \etal (1987)]{F87}Faber, S. \etal 1987, {\it Nearly Normal Galaxies}, ed. S. Faber, New York: Springer-Verlag, p.175
\bibitem[Feldman  \etal (2010)]{F10}Feldman, H. \etal 2010, MNRAS, 407, 2328
\bibitem[Forbes \etal (2011)]{F11}Forbes, D. \etal 2011, MNRAS, 413, 2943
\bibitem[Ferreras \etal (2013)]{F13}Ferreras, I. \etal 2013, MNRAS, 429, L15    
\bibitem[Graves \& Faber (2010)]{GF}Graves, G. \& Faber, S. 2010, ApJ, 717, 803
\bibitem[Hernquist (1990)]{H}Hernquist, L. 1990, ApJ, 356, 359
\bibitem[Hopkins (2013)]{Ho}Hopkins, P. 2013, MNRAS, 433, 170
\bibitem[Jones \etal (2005)]{J05}Jones, D.H. \etal 2005, PASA, 22, 277
\bibitem[Kirby et al (2011)]{K11}Kirby, E. et al 2011, ApJ, 742, L25
\bibitem[Kroupa \etal (2012)]{K12}Kroupa, P. \etal 2012, {\it Stellar Systems and Galactic Structure, Vol. V.}, arXiv 1112.3340
\bibitem[Larson (1974)]{L74}Larson, R. 1974, MNRAS, 169, 229
\bibitem[Ludlow \etal (2013)]{L13}Ludlow, A. \etal 2013, MNRAS, 432, 1103
\bibitem[Lynden-Bell \etal (1987)]{L87}Lynden-Bell, D. \etal 1987, ApJ, 326, 19 
\bibitem[McLean \etal (2007)]{M+07}McLean, I. \etal 2007, ApJ 658, 1217
\bibitem[Magoulas \etal (2012)]{M12}Magoulas, C. \etal 2012, MNRAS, 427, 245
\bibitem[Maraston  (2005)]{M07}Maraston, C. 2005, MNRAS, 362, 799
\bibitem[Mateo (1998)]{M98}Mateo, M. 1998, ARAA, 36, 435
\bibitem[Mould (1976)]{M76}Mould, J. 1976, ApJ, 207, 535	
\bibitem[Mould (1984)]{M84}Mould, J. 1984, PASP, 96, 773	
\bibitem[Mould \& Wyckoff (1978)]{M78}Mould, J. \& Wyckoff, S. 1978, MNRAS, 182, 63
\bibitem[Mould \etal (1990)]{M90}Mould, J. \etal 1990, AJ, 99, 1823
\bibitem[Navarro \etal (1996)]{NFW}Navarro, J., Frenk, C., \& White, S. 1996, ApJ, 463, 563
\bibitem[Nickerson \etal (2012)]{N}Nickerson, S. \etal 2012, ASP Conf Series, 453, 305
\bibitem[Norris \etal (2012)]{N12}Norris, M. \etal 2012, MNRAS, 421, 1485
\bibitem[Peebles (1976)]{P76}Peebles, P., 1976, Ap\&SS, 45, 3
\bibitem[Peebles (1987)]{P87}Peebles, P., 1987, Nature, 327, 210
\bibitem[Proctor \etal (2008)]{P08}Proctor, R. \etal 2008, MNRAS, 386, 1781 
\bibitem[Skrutskie \etal (2006)]{2MASS}Skrutskie, M. \etal 2006 AJ, 131, 1163
\bibitem[Smith \& Lucey (2013)]{SL}Smith, R. \& Lucey, J. 2013, MNRAS, 434, 1964
\bibitem[Smith \etal (2013)]{S13}Smith, R. \etal 2013, MNRAS, 426, 2994
\bibitem[Spiniello \etal (2012)]{S12}Spiniello, C. \etal 2012, ApJ 753, L32
\bibitem[Spinrad \& Taylor (1971)]{ST71}Spinrad, H. \& Taylor, B. 1971, ApJS, 22, 445
\bibitem[Springob \etal (2012)]{SC12}Springob, C. \etal 2012, MNRAS, 420, 2773
\bibitem[Tully \& Fisher (1977)]{TF77}Tully, R.B., \& Fisher, J. R. 1977, \aap, 54, 661
\bibitem[Weidner \etal (2013)]{W}Weidner, C. \etal 2013, MNRAS, 436, 3309
\bibitem[White (1996)]{W96}White, S. 1996, {\it Galaxy Scaling Relations}, ESO Astrophysics Symposia, eds. L. da Costa \& A. Renzini, Berlin: Springer, p.3.
\end{thebibliography}
\end{document}